\documentclass[11pt]{article}

\usepackage{mathptmx}
\usepackage{graphicx}
\usepackage{latexsym}

\usepackage{pifont}
\usepackage{booktabs}
\usepackage{amsmath,amsfonts,amssymb,nicefrac}
\usepackage{algorithm,algorithmic}
\usepackage{tabularx,multirow}
\usepackage{subfigure,color,epsfig}

\makeatletter%
\newcommand{\singlespacing}{\let\CS=
\@currsize\renewcommand{\baselinestretch}{1}\tiny\CS}
\newcommand{\singlespacingplus}{\let\CS=
\@currsize\renewcommand{\baselinestretch}{1.25}\tiny\CS}
\newcommand{\doublespacing}{\let\CS=
\@currsize\renewcommand{\baselinestretch}{1.75}\tiny\CS}
\newcommand{\extradoublespacing}{\let\CS=
\@currsize\renewcommand{\baselinestretch}{1.9}\tiny\CS}
\newcommand{\nicenicespacing}{\let\CS=
\@currsize\renewcommand{\baselinestretch}{1.9}\tiny\CS}
\newcommand{\draftspacing}{\let\CS=
\@currsize\renewcommand{\baselinestretch}{2.0}\tiny\CS}
\newcommand{\hugedraftspacing}{\let\CS=
\@currsize\renewcommand{\baselinestretch}{2.4}\tiny\CS}
\newcommand{\niceonespacing}{\let\CS=\@currsize\renewcommand{\baselinestretch}{1.1}\tiny\CS}
\newcommand{\nicetwospacing}{\let\CS=\@currsize\renewcommand{\baselinestretch}{1.2}\tiny\CS}
\newcommand{\nicethreespacing}{\let\CS=\@currsize\renewcommand{\baselinestretch}{1.3}\tiny\CS}
\newcommand{\singlespacingplusplus}{\let\CS=\@currsize\renewcommand{\baselinestretch}{1.35}\tiny\CS}
\newcommand{\nicefourspacing}{\let\CS=\@currsize\renewcommand{\baselinestretch}{1.4}\tiny\CS}
\newcommand{\nicefivespacing}{\let\CS=\@currsize\renewcommand{\baselinestretch}{1.5}\tiny\CS}
\newcommand{\nicesixspacing}{\let\CS=\@currsize\renewcommand{\baselinestretch}{1.6}\tiny\CS}
\newcommand{\nicesevenspacing}{\let\CS=\@currsize\renewcommand{\baselinestretch}{1.7}\tiny\CS}
\newcommand{\niceeightspacing}{\let\CS=\@currsize\renewcommand{\baselinestretch}{1.8}\tiny\CS}
\newcommand{\niceninespacing}{\let\CS=\@currsize\renewcommand{\baselinestretch}{1.9}\tiny\CS}
\makeatother%

\newcommand{\normalspacing}{\singlespacing}

\newcommand\qedblob{\mbox{\ding{113}}}

\newcommand{\cale}{{\cal E}}
\newcommand{\condition}{\,{\mbox{\large$|$}}\:}

  \newtheorem{theorem}{Theorem}[section]

\newcommand{\manyonereducesto}{\ensuremath{\leq_{m}^{p}}}

\newcommand{\p}{\ensuremath{\mathrm{P}}}
\newcommand{\np}{\ensuremath{\mathrm{NP}}}

\newcommand{\pspace}{\ensuremath{\mathrm{PSPACE}}}

\newcommand{\OMIT}[1]{} %

\def\literalqed{{\ \nolinebreak\hfill\mbox{\qedblob\quad}}}

\newenvironment{proofs}{\noindent{\sc Proof.}}{\literalqed\medskip}

\newcommand{\qbf}{\ensuremath{\mathrm{QBF}}}

\newcommand{\score}{\ensuremath{\mathit{score}}}

\newcommand{\onlineccac}{{\mathrm{online}\hbox{-}\cale\hbox{-}\mathrm{CCAC}}}

\newcommand{\onlineccdc}{{\mathrm{online}\hbox{-}\cale\hbox{-}\mathrm{CCDC}}}

\newcommand{\onlinesystemccac}[1]{{\mathrm{online}\hbox{-}\mathrm{{#1}}\hbox{-}\mathrm{CCAC}}}

\newcommand{\onlinesystemccdc}[1]{{\mathrm{online}\hbox{-}\mathrm{{#1}}\hbox{-}\mathrm{CCDC}}}

\newcommand{\onlinedcac}{{\mathrm{online}\hbox{-}\cale\hbox{-}\mathrm{DCAC}}}

\newcommand{\onlinedcdc}{{\mathrm{online}\hbox{-}\cale\hbox{-}\mathrm{DCDC}}}

\newcommand{\onlinesystemdcac}[1]{{\mathrm{online}\hbox{-}\mathrm{{#1}}\hbox{-}\mathrm{DCAC}}}

\newcommand{\onlinesystemdcdc}[1]{{\mathrm{online}\hbox{-}\mathrm{{#1}}\hbox{-}\mathrm{DCDC}}}

\newcounter{alg}

\newcounter{subalg}

\newcommand{\lahnote}[1]{}  \newcommand{\jrnote}[1]{} \newcommand{\ehnote}[1]{}
\newcommand{\LAHnoteHandled}[1]{}  \newcommand{\JRnoteHandled}[1]{} \newcommand{\EHnoteHandled}[1]{}

\title{The Complexity of Controlling Candidate-Sequential Elections%
\thanks{A preliminary version of parts of this paper appeared in the 
Proceedings of the 20th European Conference on Artificial 
Intelligence~\cite{hem-hem-rot:c:online-candidate-control}.
This work was supported in part by grants
ARC-DP110101792, 
DFG-RO-1202/15-1,
and NSF-CCF-\{0915792,\allowbreak{}1101452,\allowbreak{}1101479\},
and by 
COST Action IC1205,
Friedrich Wilhelm Bessel Research Awards from the 
Alexander von Humboldt Foundation,
the NRW-MIWF 
project 
``Online Partizipation,'' 
and
the SFF grant ``Cooperative Normsetting'' from HHU D{\"u}sseldorf.
This work was done
in part while E.~and L.~Hemaspaandra 
were visiting Heinrich-Heine-Universit\"{a}t D\"{u}sseldorf,
and while 
J.~Rothe was visiting the University of Rochester.}}

\author{Edith Hemaspaandra\\
        Department of Computer Science \\
        Rochester Institute of Technology \\
        Rochester, NY 14623, USA 
\and
        Lane A. Hemaspaandra\\ 
        Department of Computer Science \\
        University of Rochester \\
        Rochester, NY 14627, USA
\and
        J{\"o}rg Rothe\\
        Institut f\"ur Informatik \\
        Heinrich-Heine-Universit{\"a}t D{\"u}sseldorf  \\
        40225 D\"usseldorf, Germany
}
\date{February 29, 2012; revised June 16, 2016} 

\begin{document}
\normalspacing
\sloppy

\maketitle

\begin{abstract}  
  Candidate control of elections is the study of how adding or
  removing candidates can affect the outcome.  However, the
  traditional study of the complexity of candidate control is in the
  model in which all candidates and votes are known up front.  This
  paper develops a model for studying online control for elections
  where the structure is sequential with respect to the candidates,
  and in which the decision regarding adding and deleting must be
  irrevocably made at the moment the candidate is presented.
  We show that great complexity---PSPACE-completeness---can occur
  in this setting, but we also provide within this setting 
  polynomial-time algorithms for 
  the most important 
  of election systems, plurality.
\end{abstract}

\section{Introduction and Related Work}

This paper introduces a framework for the study 
of online candidate
control in sequential elections.  After introducing this
issue, we provide a real-world motivating example,
formalize the 
problem, provide a number of results, and suggest
directions for future work.

We will carefully define candidate control, in particular 
in our sequential 
setting, in detail later.  However, as a quick initial 
sense of what candidate control traditionally 
has meant, so that this introduction makes sense, 
candidate control refers to trying to ensure by adding or deleting 
candidates, that a given candidate wins (or does not win).  Usually,
one is limited in how many candidates one is allowed to add or delete.
Computational social choice is particularly interested in 
the algorithmic and complexity issues here: How hard it is to decide 
if in a given setting one can by such an action achieve one's goal?

Previous work on candidate control
of elections
  has been in the model of full-information, simultaneous
  voting.  This is a problem, since in quite a few real-world
  settings---from TV singing/dancing talent shows to university
  faculty-hiring processes---candidates are introduced, and appraised
  by the voters, in sequence.  We provide a natural model for
  sequential candidate evaluation, a framework for evaluating the
  computational complexity of controlling the outcome within that
  framework, and 
  results on the range such complexity can
  take on.  We hope 
this paper 
will lead to further work examining
  temporally involved candidate control, and we conclude with 
some open directions.
Our model of the process's goal, having the chair try to guarantee a
goal under the most hostile of responses, is inspired by the area of
online algorithms~\cite{bor-ely:b:online-algorithms}, and was used for online
manipulation 
in~\cite{hem-hem-rot:j:online-manipulation} 
and for 
online voter control~\cite{hem-hem-rot:c:online-voter-control}.
The just-mentioned papers~\cite{hem-hem-rot:j:online-manipulation,hem-hem-rot:c:online-voter-control}
adopt a snapshot-in-time view of 
\emph{voter-sequential} elections
(other work about or related to 
voter-sequential elections includes~\cite{slo:j:sequential-voting,dek-pic:j:sequential-voting-binary-elections,ten:c:transitive-voting,des-elk:c:sequential-voting,con-xia:c:stackelberg-sequential,par-pro:c:dynamic-social-choice}), unlike this paper, which takes 
a snapshot-in-time view with regard
to \emph{candidates} being the 
objects that are 
sequentially added.
This view is also in part inspired by the work of
Chevaleyre et 
al.~\cite{che-lan-mau-mon-xia:j:possible-winners-new-candidates-scoring},
who study the possible winner problem when new candidates are added.
Note, however, that their model and ours 
differ
greatly.
For example, while in their model addition of candidates is
not a choice (they all are just added, at once, and the question is
whether the previous votes can be extended to include this block of
new candidates such that a given candidate wins), in our model the
chair has a choice each time a new candidate shows up, and the
preferences are gradually revealed just at that moment to include
that new candidate.  Another different but related work,
done independently of and appearing 
after the preliminary version of this paper~\cite{hem-hem-rot:c:online-candidate-control},
is that of 
Oren and Lucier~\cite{luc-ore:c:budgeted-social-choice}, 
whose model 
involves votes arriving one at a time and 
in-the-moment 
choices over 
bundles of goods.
Electoral control has, in the standard (i.e., not online) setting,
been studied intensively in many papers since the seminal
work of 
{{Bartholdi}}, Tovey, and Trick~\cite{bar-tov-tri:j:control},
e.g., although this is a far from complete list,
\cite{hem-hem-rot:j:destructive-control,erd-now-rot:j:sp-av,fal-hem-hem-rot:j:llull,bau-erd-hem-hem-rot:b:approval,fal-hem-hem:j:multiprong,fal-hem-hem-rot:j:single-peaked-control,fit-hem-hem:c:control-manipulation,men:j:range-voting,rot-sch:j:typical-case-challenges,erd-hem-hem:c:vru-bribery,fal-hem-hem:j:single-peaked-nearly,erd-fel-rot-sch:j:control-in-bucklin-and-fallback-voting,erd-hem-hem:c:more-natural-models-of-partition-control},
see also the detailed 
survey~\cite{fal-rot:b:handbook-comsoc-control-and-bribery}.

\section{Motivating Example}
\label{app:motivating-example}
In an author's school, faculty hiring happens basically as follows.
On some Mondays, a candidate visits, gives a talk, and meets with
faculty members.  Then each of the department's rank-and-file faculty
members sends by email to the faculty and department chair 
her ranking of all the
candidates so far, namely, by inserting the new candidate into the
preference order she sent after the previous candidate.  The
chair typically follows up 
by phoning the
candidate a day or two after the visit, so that phoning occurs after the chair has
seen the faculty rankings generated by the candidate's visit.  Moving
now from reality to (slight?)~fiction, let us imagine that the chair
in that followup can easily choose to scare away a candidate (``Oh,
did I remember to mention that if you come,
your office will be a
shared closet in our lovely basement, I'll help you broaden yourself by
teaching a wide range of introductory courses, and I see in you a real
talent for extensive committee work which I'll put to good use?'').
But let us further assume that the chair cannot do this more often
than a certain threshold, as otherwise the rank-and-file faculty will
realize the chair is manipulating the process and will revolt.  So,
how should the chair use this power of candidate suppression to most
effectively ensure that one of the candidates the chair likes will, at
the end of the process, win the election (under the faculty preferences,
among the candidates not scared away)?

This example nearly perfectly captures the topic and model of
this appendix.  We are moving what in the literature is called
``candidate control''~\cite{bar-tov-tri:j:control} (in the example, of
the sort known as ``constructive control by deleting candidates'')
from its existing setting of simultaneous elections into a setting
where preferences are set/revealed sequentially and the chair, right
after the preferences related to an introduced candidate are revealed, must
use-or-forever-lose the ability to suppress that candidate.  

We also are interested---again moved to 
a sequential setting---in constructive control by adding candidates, a
natural analogue of the above, and in destructive versions of both
adding/deleting candidates, which are the same issues except the
chair's goal is to ensure that none of a certain set of hated
candidates is hired.

Bartholdi, Tovey, and Trick~\cite{bar-tov-tri:j:control}
defined non-online versions of the
constructive-deletion notion used above and a precursor of the
constructive-addition notion used above.  The non-online versions of
the constructive-addition notion used above and both destructive
notions used above are from Hemaspaandra, 
Hemaspaandra, and Rothe~\cite{hem-hem-rot:j:destructive-control},
although destructivity had been introduced even earlier by Conitzer,
Sandholm, and Lang~\cite{con-lan-san:j:when-hard-to-manipulate} for a
different type of attack known as manipulation.  (However,
as mentioned earlier in the paper, we are consistently following 
the now more standard nonunique-winner model, rather than the 
unique-winner model.)

\section{Formalizing the Problem}
Let us discuss how to formalize this into a decision problem whose
complexity can be studied.  
We'll do so here in 
detail just for
constructive control by deleting candidates, and then 
will describe, by altering that, how the other online
candidate-control problems are captured.
Let $\cale$
denote the underlying election system:
a mapping from candidates and
votes over the candidates (with preferences typically as strict,
linear orderings)
to a set of winners.  The candidates
left standing at the end (i.e., not deleted by the chair)
will be fed into this election system along with the 
votes (with each vote's preference order masked down to
that set of still-standing candidates).

The input will capture a
``moment of decision''\label{words:moment-of-decision} for the chair.  That
is, the input will give the history of the process up to the given
point, and then will ask whether there is some action of the chair that can
ensure she will get a happy outcome.  We must make it clear what we
mean by this.

The input will be the set of candidates, the set of voters, the order
in which the candidates will be presented, a flag denoting which the
current candidate is, a bound $k$ on the maximum number of candidates
the chair can suppress, an ordering $\sigma$ of how the chair views all
candidates (to put this in the context of 
our motivating example from Section~\ref{app:motivating-example},
this is as if the department chair
had the c.v.'s ahead of time and has
evaluated them already), a specific candidate $d$ such that the
chair's goal is to ensure that there is an election winner from the
set $\{c \condition c \geq_\sigma d\}$ (i.e., $d$ or some candidate
the chair likes better than $d$ is a winner), and the history up to
the current moment in time (which means for each candidate before the
current one a bit saying whether the chair deleted that candidate, and
a preference order for each voter over all the candidates up to and
including the current one---we could also make this just over all
as-yet nondeleted candidates, but let us make it over all candidates
so far, though it doesn't affect the eventual results; we prefer
this because
it allows 
the history of the voting situation to be part of the instance).  And the
question being asked in this decision problem is whether there is some
decision the chair can make about the current candidate (to delete, or
not to delete) such that, assuming that the chair at each future
decision is free to act in light of the information revealed up to
that point, the chair can ensure that the winner set will have
nonempty intersection with the candidates she likes, 
$\{c \condition c \geq_\sigma d\}$, regardless of what else happens in the
election (i.e., even if the revealed preferences are highly
unfavorable to the chair's wishes).

The decision problem (i.e., language) here is simply the set of all
inputs where the answer to that question is yes.  Let us call this
problem \emph{online-$\cale$-constructive-control-by-deleting-candidates}
($\onlinesystemccdc{\cale}$, for short).

Although we used a somewhat 
informal wording above, there is a more formally
satisfying phrasing
that captures the same notion using
alternating quantifiers: Does there exist
a legal move by the chair about the current candidate, such that
for all possible settings of the information revealed after this up to the
chair's next decision, there exists a legal next decision by the chair,
such that $\ldots$~$\ldots$~such that the winner set contains either $d$ or some
candidate the chair likes more than $d$.  

Briefly, the ``adding''-candidates analogue (of the 
above deleting-candidates case), denoted by
$\onlinesystemccac{\cale}$, is almost the same---except
the input contains a ``certainly in the election'' set of candidates,
and a (disjoint) 
set of ``potential additional'' candidates, and a presentation
ordering over the union of those two sets, and the rest is analogous (so for
potential-addition candidates before the current one the input tells
whether the chair added them, etc.).

And these constructive-control deleting and adding cases each have a
``destructive control'' sibling, $\onlinesystemdcdc{\cale}$ and
$\onlinesystemdcac{\cale}$, where the question is whether the
chair can ensure that no one ``$d$ or worse'' is a winner
(i.e., the chair can ensure that
no member of 
$\{c \condition d \geq_{\sigma} c\}$ is a winner).  

For
destructive control by deleting candidates, there is a special issue
as to whether the chair can simply start deleting some or all
candidates who are ``$d$ or worse,'' thus perhaps ruthlessly obtaining
her goal.  Our default model---call it the ``non-hand-tied chair''
model---is that the
chair may delete some, but never all, of the candidates who are ``$d$
or worse.''  An alternative model---call it the ``hand-tied chair'' model---is
that the chair may never delete anyone who is ``$d$ or worse.''  The
results we mention in this appendix for destructive control by deleting
candidates hold equally well for both those models.  In both these 
models, in any legal 
input instance,
the ``previous actions'' by the chair
\LAHnoteHandled{I added ``by the chair,'' as otherwise this might 
make some people wonder (though they should not)
whether all previous votes-fillings-in had to 
be opposed to the chair, but that is just the look-forward 
pessimistic side of our definitions, and it does not 
bind the PAST (just happened) votes that have been revealed, which can
by good fortune be not hostile to the chair.}%
cannot violate the model, e.g., in the hand-tied chair model,
if the history shows that some candidate in 
$\{c \condition d \geq_{\sigma} c\}$ was deleted, then the input is rejected,
as it is illegal.  The history the
snapshot provides can legally contain dumb actions, but it cannot
contain illegal ones.

As always, in 
the language of multiagent systems candidates are alternatives and
voters are agents.  So though about ``elections,'' this model is
equally well about preference aggregation in multiagent systems in
which the alternatives are sequentially revealed and evaluated by the
agents, and another party is trying to control the outcome.

\section{Complexity Results}\label{s:appendix-candidate}

Let us assume that our election system's ($\cale$'s)
winner-determination problem (i.e., ``Is candidate $c$ a winner under
this election system,
if the candidates and votes are $C$ and
$V$?'') is in polynomial time.  Then it is easy to see 
{}from the quantifier approach mentioned above
that all our
above 
online candidate control problems 
are in PSPACE\@.
The PSPACE upper bound remains valid 
even if we restrict $\cale$'s winner problem not to P but rather 
to PSPACE\@.

Clearly, not all election systems will require the full power of
PSPACE for mounting control attacks.  It is easy to construct
artificial systems where all these control attacks have
polynomial-time 
algorithms.
But a more important question
is whether the PSPACE upper bound is itself too enormous.  Can 
such tremendous control complexity be realized, even for election
systems whose winner problems must be in polynomial time?

The answer is yes.  Although the construction is not simple, we have
by setting up appropriate election systems and reductions from
intractable problems, shown that for each of the
problems defined
above, there is an election 
system with a polynomial-time winner problem for which 
the online control problem 
of the given type is PSPACE-complete.

Briefly put, the construction enmeshes issues of formulas into
election systems in a way that so tightly incorporates 
and interprets formulas,
variables, and assignments, that one can---by using a careful reduction
and some legal preprocessing transformations---ensure that the process
of the online control attempt can succeed exactly if the input to a 
PSPACE-complete formula-problem that transformed into that problem is
a positive instance.

\begin{theorem}
\label{thm:online-candidate-control-general-pspace-upper-lower-bounds}  
\begin{enumerate}
\item  For each election system $\cale$ with a polynomial-time winner
problem,\footnote{The first statement of
  Theorem~\ref{thm:online-candidate-control-general-pspace-upper-lower-bounds}
  holds even for election systems whose winner problems are in
  $\pspace$.}
$\onlineccdc$, $\onlineccac$, $\onlinedcdc$ (in both the
  non-hand-tied and the hand-tied chair model), and $\onlinedcac$
  are in $\pspace$.

\item There exist election systems $\cale$ and $\cale'$ with
  polynomial-time winner problems such that $\onlineccdc$,
  $\onlineccac$, $\onlinesystemdcdc{\cale'}$ (in both the
  non-hand-tied and the hand-tied chair model), and
  $\onlinesystemdcac{\cale'}$ are $\pspace$-complete.
\end{enumerate}
\end{theorem}

\begin{proofs}
1. 
The
famous characterization of PSPACE as alternating 
polynomial time, due to Chandra, Kozen, and 
Stockmeyer~\cite{cha-koz-sto:j:alternation}, establishes the upper bounds for
these four problems: Each can be solved by an alternating Turing
machine in polynomial time, and thus by a deterministic
polynomial-space Turing machine.

2. 
\LAHnoteHandled{This proof is newly 
(namely, in 3/2015) added by JoergR, from Lane's notes I think.
So it has not been proofread (except of course by Joerg).  Edith or 
Lane should do a careful proofreading of it, since it is new to 
the paper, and thus is one of the most unchecked things in the paper.}%
\LAHnoteHandled{Ok... on 151129 I did a (relatively quick) read-through
and very slight re-editing, but the proof seems fine (at least relative
to doing just a relatively quick read-through).}%
For proving the lower bounds, we will define
$\manyonereducesto$-reductions from the $\pspace$-complete problem
$\qbf$ to our online candidate-control problems.  In fact, we will
prove that these problems are PSPACE-complete even when limited 
to the case of there being one voter.

We start by providing the $\manyonereducesto$-reduction from $\qbf$ to
$\onlineccac$ for the election system $\cale$ defined as follows.
Interpret each candidate as a pair $(F,i)$, where $F$ is a boolean
formula and $i$ a nonnegative integer.  If there are any syntactic
problems, or if any two candidates have \emph{distinct} boolean
formulas, then everyone loses.  Otherwise, all candidates have the
same boolean formula, call it~$\hat{F}$.  Let
\JRnoteHandled{I rename ``$k$'' to ``$\ell$'' because we usually use $k$ to
  denote the addition/deletion bound.}
$\ell$ be the number of
variables in $\hat{F}$ (e.g., $\hat{F} = (x_1 \vee \neg x_2)
\Longleftrightarrow (x_3 \wedge \neg x_3 \wedge x_3 )$ has three
variables: $x_1$, $x_2$, and $x_3$).  We assume that $\ell \geq 1$
(otherwise, $\hat{F}$ is syntactically illegal, so everyone loses).
Now, if
(a)~$\ell$ is odd,\footnote{Our reductions will always map to
  formulas of the form $(\exists w_1)\, (\forall w_2)\, \cdots
  (\exists w_{2j-1})\, (\forall w_{2j})\, [\cdots ]$.}
  or
(b)~the candidate set does not contain $(\hat{F},i)$ for every even~$i$,
  $0 \leq i \leq \ell$,
  or
(c)~there are two or more voters,
then everyone loses.  Otherwise, lexicographically order the $\ell$
variables by their names.  We will refer to the lexicographically
$i$th among them as $v_i$ for the purpose of this proof.  For each
odd~$i$, $1 \leq i \leq \ell$, set $v_i$ to true if and only if there is
a candidate named $(\hat{F},i)$.  For each even~$i$, $2 \leq i \leq
\ell$, set $v_i$ to true if and only if the (single) voter prefers
candidate $(\hat{F},i)$ to candidate $(\hat{F},0)$.  Now, if $\hat{F}$
is true under this assignment then everyone wins; else everyone loses.
This ends the specification of election system~$\cale$.
Note that the winner problem for $\cale$ is in~$\p$.

We now $\manyonereducesto$-reduce $\qbf$ to $\onlineccac$.  Let $G$ be
a given $\qbf$ instance, i.e., we want to know whether $G$ belongs to
$\qbf$.  Without loss of generality, let $G$ be of the form $(\exists
w_1)\, (\forall w_2)\, \cdots (\exists w_{2j-1})\, (\forall w_{2j})\,
[\varphi(w_1, w_2, \ldots , w_{2j})]$, where $j \geq 1$ and $\varphi$
is a propositional formula.  Now rewrite $G$ and $\varphi$ so that their
actual variable names are lexicographically ordered in the order $w_1,
w_2, \ldots , w_{2j}$.  Then remove all quantifiers.  Call what that
creates~$\hat{F}$; it is basically $\varphi$ with variable names
adjusted as above.  

We will map this to the following instance of 
$\onlineccac$:
\begin{itemize}
\item the initial set 
of already qualified candidates is 
$\{(\hat{F},0), \allowbreak (\hat{F},2), \ldots ,\allowbreak
  (\hat{F},2j)\}$;
\item the set 
  of spoiler candidates that can potentially be added is
$\{(\hat{F},1), (\hat{F},3), \ldots , (\hat{F},2j-1)\}$;
\item the addition bound $k$ is $j$;
\JRnoteHandled{I added the addition bound~$k = j$.}%
\item the chair's preference order~$\sigma$ is $(\hat{F},2j) >_{\sigma}
  (\hat{F},2j-1) >_{\sigma} \cdots >_{\sigma} (\hat{F},0)$;
\item the distinguished candidate $d$ is $(\hat{F},0)$;
\item the presentation order 
of the candidates 
is 
$(\hat{F},0), (\hat{F},1), \ldots ,
  (\hat{F},2j)$;
\item the 
  current candidate, $c$, for whom the chair has to make a decision now
  as to whether to add her is $(\hat{F},1)$; and
\item the voter set is a single voter
whose preference with regard to $(\hat{F},0)$ and $(\hat{F},1)$ is
(actually irrelevant for the reduction but we are required by our
model to provide it): $(\hat{F},0)$ is preferred to $(\hat{F},1)$.
\end{itemize}
This completes the description of the $\manyonereducesto$-reduction
from $\qbf$ to $\onlineccac$,
which clearly is computable in polynomial time.  It remains to prove
that it is correct.  This, however, is easy to see: By definition of
the election system~$\cale$, $G \in \qbf$ if and only if the chair can
make some decision about the current candidate $c$ (to add, or not to
add) such that, assuming that the chair at each future decision is
free to act in light of the information revealed up to that point, the
chair can ensure that the winner set will have a nonempty intersection
with the candidates she likes, $\{(\hat{F},i) \condition (\hat{F},i)
\geq_\sigma (\hat{F},0)\}$ (which happens to be all
candidates),\footnote{That is, the question of $\onlineccac$ 
will in this construction boil down
  to whether at least one candidate wins.  Recall that in $\cale$
  either everyone wins or everyone loses.}
regardless of what else happens in the election.

$\pspace$-completeness of $\onlineccdc$ is proven very similarly.  We
use the same election system, $\cale$, and from a given $\qbf$
instance $G$ (as above) we construct an $\onlineccdc$ instance, where
we start with the candidate set $\{(\hat{F},i) \condition 0 \leq i
\leq 2j\}$ and now set the \emph{deletion} bound to $k=j$.  Everything
else in the reduction remains the same.  Again, it follows that $G \in
\qbf$ if and only if the chair can make some decision about the
current candidate $c$ (to delete, or not to delete) such that,
assuming that the chair at each future decision is free to act in
light of the information revealed up to that point, the chair can
ensure that at least one candidate (all of which the chair likes)
wins, regardless of what else happens in the election.

The destructive cases can be handled quite similarly.  The only
difference is that we define election system $\cale'$ to be just like
$\cale$ except that we change every occurrence of ``everyone loses'' to
``everyone wins'' and every occurrence of ``everyone wins'' to
``everyone loses.''  In our reductions from $\qbf$ to the destructive
control problems, using the same $\sigma$ as the chair's preference
order and the same distinguished candidate $d = (\hat{F},0)$ will be
fine.  In particular, it will not conflict with either the
non-hand-tied or the hand-tied chair model: Under each of those
models, specifying this $d$ and this $\sigma$ means ``we cannot delete
candidate $(\hat{F},0)$,'' so deleting $(\hat{F},0)$ would be illegal;
therefore, to keep this candidate from winning, the only way is to
ensure the ``everyone loses'' triggers due to $(\exists w_1)\,
(\forall w_2)\, \cdots (\exists w_{2j-1})\, (\forall w_{2j})\,
[\hat{F}(w_1, w_2, \ldots , w_{2j})]$ holding.  But this implies for
each considered case of destructive online candidate control
($\onlinedcdc$ in both the non-hand-tied and the hand-tied chair model
and $\onlinedcac$) that the given control instance is positive if and
only if $G \in \qbf$.~\end{proofs}

We now turn to online candidate control for some specific election
system widely in use: plurality.  The following result shows that both
constructive and destructive online control by adding and deleting
candidates is an easy problem for (candidate-sequential) plurality
voting, in sharp contrast with the corresponding 
standard 
control problems, which are $\np$-complete for
plurality (for the two 
constructive cases, 
this is due to the fact that 
the two relevant unique-winner-model results of 
\cite{bar-tov-tri:j:control,hem-hem-rot:j:destructive-control}
have been verified 
in \cite{fal-hem-hem:j:single-peaked-nearly}
to also hold in the 
nonunique-winner model;
for the two destructive cases, this is due to the fact 
that we have verified that the two relevant unique-winner-model results of 
\cite{hem-hem-rot:j:destructive-control}
also hold in the 
nonunique-winner model).
\LAHnoteHandled{It says ``we have verified'' but we in fact have not yet 
done so, though doing so should be easy.  Joerg, would you please do that 
quick verification, send us an email that you've checked it, and then we can 
embed your email and its proof/explanation 
in this paper as an invisible comment to 
document that it has been checked and holds (see for example 
my own doing of that earlier in the source for plurality-DCDV-NUW and 
plurality-DCAV-NUW), namely, that 
(non-online) plurality-DCDC-NUW  and (non-online) plurality-DCAC-NUW 
are NP-complete?}%
\EHnoteHandled{The UW vs.\ NUW issue at THIS (and thus at every)
location has been handled.
I have checked and embedded into our source code,
as invisible comments, Joerg's explanation 
of both cases that are not previously stated in the literature.}%

\begin{theorem}
\label{thm:online-ccdc-plurality}
$\onlinesystemccdc{\mathrm{plurality}}$,
$\onlinesystemccac{\mathrm{plurality}}$,
$\onlinesystemdcdc{\mathrm{plurality}}$ (in both the non-hand-tied and
the hand-tied chair model), and
$\onlinesystemdcac{\mathrm{plurality}}$ are
in~$\p$.
\end{theorem}

\begin{proofs}
  Consider an input to the problem
  $\onlinesystemccdc{\text{plurality}}$ as defined above.
  So we are focused on some 
  current ``moment of decision'' for the chair (recall what this means
  from page~\pageref{words:moment-of-decision}).
We
describe a polynomial-time algorithm for the question:
Does the 
chair have a current action 
that will ensure her of reaching her goal?

Let $d$ be the distinguished candidate and $\sigma$ be the chair's
preference.  We refer to the candidates in
$\{a \condition a \geq_{\sigma} d\}$ as \emph{good} candidates
and to the other ones as \emph{bad} candidates.
As required in this moment of decision
described on page~\pageref{words:moment-of-decision}, each of the
already revealed candidates have their flags set as to whether or not
they have been deleted, and we assume that the votes are currently
masked down to the \emph{still-standing candidates} (i.e., to the
already revealed, yet not deleted candidates up to this point in time).

Rather than analyze directly what to do in this moment, let us note
that we have at most two choices in this moment.  Either we leave $c$
in, or (only if the number of allowed deletions has not been already
expended with the already done deletions) we remove $c$.  Each of those
cases leaves a relatively pure situation, in which $c$ is no longer
special.  And so what we will now do is show how to analyze such a pure
situation, i.e., to say, in such a situation, whether the chair has a
forced win.  (Knowing how to do that in polynomial time implicitly
resolves our problem.  We just check the two cases and see if at 
least one is a
forced win for the chair---except if the deletion
bound was already expended then we
check only the case where $c$ is kept in and we see if that is a forced-win
setting.)

So, what does such a case look like?  The setting is we now are given:
the set of candidates; an ordering $\sigma$ over the candidates; a
designated candidate $d$ (recall: our goal is to ensure that 
the winner set has nonempty intersection with the set 
$\{a \condition a \geq_\sigma d\}$);
an order $\tau$ in which the candidates are being revealed, where
the candidates in a nonempty
prefix of $\tau$ (i.e., the ones who have had the preferences
among them revealed so far) are flagged as to being either
removed or kept in (and all candidates after that in $\tau$ are not
yet flagged as being in or
out);\footnote{The prefix of $\tau$ is nonempty because ``$c$''
  will already have been revealed, since we call this polynomial-time
  routine only, see the previous paragraph, when at least one
  candidate has been set as in-or-out.}
all votes but
masked down to just the still-standing candidates (the already
revealed but not deleted candidates); and 
a natural
number $k \geq 0$,
which will be how many deletions are left to
use.\footnote{If from the prefix of $\tau$ and the flags we find more
  deletions have happened than the problem originally allowed, a case
  that won't actually ever happen within the way we are calling this
  algorithm, we are then already in a no-win case since we, the chair,
  back in the given history already violated the deletion bound.  So
  if this (impossibly) were to happen, then we would return the fact that
  forced-success along a legal path from our current state is
  impossible, since we've already violated the rules.}

And we need to know whether we have a forced win even if the universe
is perfectly hostile to us, i.e., we need to know whether 
no matter what votes are revealed
as the process progresses we will win (i.e., the winner set will have
a nonempty intersection with $\{a \condition a \geq_\sigma d\}$).

Let $b$ be the number of bad candidates currently unrevealed.

Our polynomial-time algorithm proceeds as follows.  If there is no
voter, every candidate that is left standing
at the end will be a plurality winner (with score zero).
So in this case our algorithm accepts if there is at least one 
good candidate revealed but undeleted, or not yet revealed.
And otherwise our algorithm rejects.  So henceforward, let us 
assume that the number of voters is at least one.

If all candidates are already revealed, just analyze whether a 
good candidate is among the winners, and if so, return success on
this case.  Otherwise, we go on as follows.

We now check whether
any good candidates are currently winning.
If no good candidate wins at the current moment, then 
(i)~if there is
at least one unrevealed good candidate and the number $b$
\JRnoteHandled{I added ``$b$''.}%
of unrevealed
bad candidates is at most $k$, then we have a forced win here; 
(ii)~otherwise 
return that we don't have a forced win here (for example,
we certainly won't win if 
the votes will be eventually revealed to show that the future---i.e.,
as yet unrevealed---bad
candidates are a top segment of every voter's vote).

\LAHnoteHandled{The paragraph following this duplicates an
  argument-flavor/reasoning of the last part of the previous
  paragraph, and probably one could merge them to have this more
  concise...  but I've simply said ``much as above'' below to make
  clear to the reader that we're aware of the repetition of the
  argument type, and have left things at that, as that is fine.}

Otherwise, at least one good candidate is currently winning (according
to the votes masked down to the still-standing candidates up to the
current point).  We first check whether $k \geq b$, i.e., whether all
bad candidates after $c$ can be deleted.  If this is not the case,
then much as above there is no hope for the chair to be sure to
reach her goal, since for example even a single future bad candidate that cannot
be deleted may be ranked on top of every vote in the worst case.

Let us finally consider (with all the things eliminated above 
\JRnoteHandled{I'm not sure if ``still being handled'' is right.  Is perhaps
  ``already been handled'' better?  Or can it just be deleted?; NOTE ADDED BY LANE: rewritten}%
not being on the table as possibilities that we 
have to deal with here) 
the case where at least one
good candidate is currently winning and $k \geq b$.  If there 
are no 
revealed,
\JRnoteHandled{I was going to ask why ``[sic]'' is there; now I see the answer
  in the latex source.  I would remove ``[sic]'' because it might puzzle
  the reader (who doesn't see the latex source).}%
still-standing
bad candidates, return that we have a forced win (as there will
be no bad candidates at the end and at least one good one, so 
there will certainly be a good one among the winners),
and so henceforward assume we have at least one revealed,
still-standing bad candidate.

Let $B$ be the highest current plurality score among all revealed,
still-standing
bad candidates,
according to the votes masked down to the revealed, still-standing
candidates.

A best strategy for the universe,
regarding seeking to defeat the chair's goals, 
is as follows.

All future bad candidates 
are in a top-segment of
each vote.  This forces the chair to expend $k-b$ deletions on
removing those future bad candidates.

The interesting twist here is what a hostile universe can do with
future \emph{good} candidates.  Note that future good candidates can,
if not deleted, steal first-place votes from existing good candidates
that are winning or otherwise doing well.  

The most effective attack a hostile universe can do, given that it is
already going to force us to remove all future bad candidates, is to
leave $B$ as the strongest score of any bad candidate (the universe
has no tool left to increase beyond $B$ since it is---for 
a different compelling reason---making us spend our deletions 
to burn away all
future bad candidates, and the existing bad candidates already are revealed and
so can only move down in how many votes they are at the top of as more 
candidates are
revealed in the votes; but moving them down would play against the universe's
interests here).  So the universe's best strategy is to 
also 
use new good candidates to try to
decrease the scores of all revealed, still-standing good candidates to
at most $B-1$, while also ensuring that no added 
good candidates score more than $B-1$.  
Of course, the chair would try to stop this, by
deleting good candidates as needed (at most $k-b$ deletions of 
good candidates though, on top of the $b$ deletions of bad candidates
that the universe can force to be needed bad-candidate deletions).

Let $g$ be the number of future good candidates.  If $k-b \geq g$, we
(the chair) have a forced win
(recall that a good candidate is currently winning), 
as we can remove all of those
candidates.  

What about the case $ k-b < g$?  Note that the universe then can add
$g - (k-b)$ good candidates (of our choice, not its choice, but it
turns out that our choice won't help us here against an optimally
hostile universe).  The most effective thing the universe can try to
do is to use
\JRnoteHandled{I changed ``the new good candidates'' to ``these new good candidates''
  to specifically refer to the above-mentioned
  $g - (k-b)$ good candidates we cannot delete.}%
these new good candidates to try to draw all the
revealed, still-standing
\JRnoteHandled{I changed ``revealed-still-standing'' to ``revealed, still-standing''.}%
good candidates down to at most $B-1$ first-place
votes, while also not letting any still-standing-at-the-end 
good candidate (including ones added after the moment this problem
is looking at) get more than $B-1$ first-place votes.  

Let $S$ denote the set of all revealed, still-standing
candidates $i$ for which $\score(i)\geq B$.
\JRnoteHandled{I changed ``$\score(i)\geq b$'' to ``$\score(i)\geq B$''.}%
Note that the ``surplus'' of each member $i$ of $S$ is  
$\score(i) - (B-1)$, and the universe is trying to in effect 
shed these surpluses onto future good candidates that don't get 
removed by the chair, while ensuring
that none of those candidates itself ends up with $B$ or more 
first-place votes.  The universe will have at least $\|S\| +
g - (k-b)$ candidates over which to do this spreading.  

It is thus not hard  to see that if
$$\left\lceil 
\frac{\sum_{i \in S}
\score(i) 
}
{\|S\| + g - (k-b)}
\right\rceil \geq B,
$$
\JRnoteHandled{In the above inequality, I've changed
  ``$\sum_{\{i ~|~ i \in S\}}\score(i) $'' to
  ``$\sum_{i \in S}\score(i) $.''}%
then we (the chair) 
have a forced win, since we can ensure that at least one current
or added good candidate will have score at least $B$.  And if the
above ceiling is less than $B$, we do not have a forced win, as the
universe can by adding good candidates ensure that all good candidates
still-standing at the end have score at most $B-1$.\footnote{How does the
universe do this, given that it doesn't know which up to $k-b$ good
candidates we will delete?  If it did know which, how to do this
balancing is immediate.  
But what it will do is it will as the votes
are revealed for each future good candidate make them consistent 
on the up-to-then revealed candidates with
what it would do with those votes if that one were to be the next
member of its set of at least $g - (k-b)$ good candidates to add.
If the chair removes the candidate, it will then try that same 
approach again with the next to-be-revealed good candidate.  
Not knowing which still-in-the-future-at-that-point candidates
will be allowed in is not a problem, since the universe does not have 
to commit to their locations in the votes 
until the moment the candidate is revealed
for possible deleting by the chair.
So by the end, the universe has indeed added a set of at least $g  - (k-b)$
good candidates such that no still-standing good candidate has 
more than $B-1$ first-place votes.
Note of course that this process 
never requires 
the universe to have
a member of $S$ take a point ``away'' from another member of 
$S$, which is good as that simply can't be done.  Rather, it is the incoming 
good votes that are used to drain away the points from the members
of $S$.}
\JRnoteHandled{In the previous footnote, I changed the first occurrence of
  ``$g - (k-b)$'' (namely, in ``which up to $g - (k-b)$ good candidates
  we will delete'' to ``$k-b$''.  I also removed the hyphens in
  ``locations-in-the-votes''.}

This completes our polynomial-time algorithm (that is ``called'' up
to two times---with $c$ left in, and if $k>0$ then also with 
$c$ removed), and so gives us an overall polynomial-time 
algorithm for 
$\onlinesystemccdc{\text{plurality}}$, as promised.

\LAHnoteHandled{And, joerg, if the above isn't wrong, please then 
fill in below the handwaving arguments to state that all the 
other cases hold.  References to things like what formerly
were called (i) (ii) (a) (b) and so on won't work, since those are 
not there.  of course, you can as needed add numbering above, but 
try not to change the content (unless it is outright wrong), since
small changes can easily added errors and missed cases.

do be careful.  for example, when handwaving destructive cases, the
case of not having any good/bad votes from a given point on or at a
given point, which the above is handling quite carefully as a
mentioned special case (and the previous proof was often mishandling)
may not be perfectly symmetric, and so and overly detailed handwave
could make assertions that---like certain in each previous
proof---were simply clearly incorrect.  so perhaps best is to very,
very vaguely handwave, essentially by just asserting that all the
other cases similarly hold.

\emph{NOTE TO JOERG: And still left in the paper, as TINY text,
is all the original text regarding the 
other cases, in case you want to edit it to now cover the remaining 
cases, except rewritten to sync with the above proof.}
}%

That $\onlinesystemccac{\text{plurality}}$ is in $\p$ can be shown
similarly.  A difference is that now we have two types of candidates:
\emph{qualified} candidates who are certainly in the election and
\emph{spoiler} candidates who can possibly be added by the chair.
Therefore, instead of speaking of candidates to be deleted (or not),
our algorithm will now speak of spoiler candidates to be added (or
not).  In particular, when we are in the case that at least one good
candidate is currently winning (according to the votes masked down
to the still-standing candidates up to the current point), in order
to check whether no bad candidate after $c$ can enter the election
(which would immediately destroy the chair's hopes), we now need to
check whether all bad candidates after $c$ are spoiler candidates
(which the chair simply doesn't add).\footnote{Note
  that the chair can decide \emph{not} to add any of the future
  spoiler candidates; only the number of spoiler candidates that can
  be added is limited.  In fact, the chair will never add a bad
  spoiler candidate coming after~$c$.  Note also that just a single
  future candidate that is both qualified and bad would kill off the
  chair's chances to reach her goal, since this candidate will take
  the top position of each vote in the worst case.}
Now, the hostile universe's best strategy against the chair (leading
to our worst-case scenario) can be described similarly to the above
best strategy of the universe in the case of
$\onlinesystemccdc{\text{plurality}}$, except that we consider ``good
candidates after $c$ that are qualified'' instead of ``good candidates
after $c$ that cannot be deleted.''

The destructive cases can be handled analogously as well, with just a
few
changes.
Consider $\onlinesystemdcdc{\text{plurality}}$ with essentially the
same notation used above for $\onlinesystemccdc{\text{plurality}}$.
In particular, $d$, $\sigma$, $c$, $k$, $b$, and $B$ have the same
meaning, except that the distinguished candidate $d$ has now turned
from a good into a bad candidate, since the chair's goal now is to
make sure that no one ``$d$ or worse'' is a winner (i.e., we now refer
to the candidates in $\{a \condition a >_{\sigma} d\}$ as \emph{good}
candidates and to the other ones as \emph{bad} candidates).  We can
handle both the \emph{non-hand-tied chair model} (where, recall, the
chair may delete some, but never all, bad candidates) and the
\emph{hand-tied chair model} (where, recall, the chair may never
delete any bad candidate).  However, here we will simply mention 
just some key differences, starting with the former.
For instance, when we are in the case that at least one good candidate
is currently winning (according to the votes masked down to the
still-standing candidates up to the current point) and we need to
check whether all bad candidates after $c$ can be deleted in the
non-hand-tied chair model, we have to check whether either some
already revealed bad candidate up to now has been labeled as undeleted
and $k \geq b$, or no already revealed bad candidate up to now has
been labeled as undeleted and $b=0$.  If this is not the case, there
is no hope for the chair to reach her goal, as even a single future
bad candidate that cannot be deleted in the non-hand-tied chair model
will be ranked on top of every vote in the worst case and thus wins.
That is, the chair doesn't have a forced win in this case.  Similarly,
the rest of the argumentation can be slightly adapted to show that
$\onlinesystemdcdc{\text{plurality}}$ in the non-hand-tied chair model
is in~$\p$.

The proof that in the hand-tied chair model also 
$\onlinesystemdcdc{\text{plurality}}$ is in $\p$ differs from the
above proof only slightly.  For instance, instead of checking whether
all bad candidates showing up after $c$ can be deleted, we now 
check only whether $b=0$, as we are not allowed to delete any bad candidate
in the hand-tied chair model.

Finally, incorporating in the proof of
$\onlinesystemccac{\text{plurality}} \in \p$ the changes corresponding
to those that turned the proof of $\onlinesystemccdc{\text{plurality}}
\in \p$ into a proof of $\onlinesystemdcdc{\text{plurality}} \in \p$
(though disregarding the issue of whether the chair is hand-tied or
not), we see that $\onlinesystemdcac{\text{plurality}}$ is in~$\p$ as
well.~\end{proofs}

\section{Conclusions and Open Directions}
This paper's contribution is a model and a number 
of results for the research 
direction of candidate-sequential
elections---a direction that we feel 
is of interest, not as a replacement for 
the study of voter-sequential elections, but as a notion that captures 
different but also important settings.  
Our results show that online candidate-control can 
be extremely complex, but that for the most important
real-world election system, plurality, candidate control can be 
quite simple---even 
of polynomial
time-complexity.

It will be important to seek
further results for the complexity, in this
model, of natural systems.
It would also be interesting to formalize and study online control by 
partition of candidates in sequential elections.
Another interesting direction will be to
also give the chair limited or total control over the candidate
presentation order; in political science, for example, in many settings
control of agenda-order can be powerful.

\section*{Acknowledgments}
Parts of this paper appeared in preliminary form in 
ECAI-2012~\cite{hem-hem-rot:c:online-candidate-control}, and
we are very grateful to the 
reviewers for their comments and suggestions.

\bibliographystyle{alpha}

\begin{thebibliography}{CLM{\etalchar{+}}12}

\bibitem[BE98]{bor-ely:b:online-algorithms}
A.~Borodin and R.~{El-Yaniv}.
\newblock {\em Online Computation and Competitive Analysis}.
\newblock Cambridge University Press, 1998.

\bibitem[BEH{\etalchar{+}}10]{bau-erd-hem-hem-rot:b:approval}
D.~Baumeister, G.~Erd\'{e}lyi, E.~Hemaspaandra, L.~Hemaspaandra, and J.~Rothe.
\newblock Computational aspects of approval voting.
\newblock In J.~Laslier and M.~Sanver, editors, {\em Handbook on Approval
  Voting}, pages 199--251. Springer, 2010.

\bibitem[BTT92]{bar-tov-tri:j:control}
J.~{{Bartholdi}}, III, C.~Tovey, and M.~Trick.
\newblock How hard is it to control an election?
\newblock {\em Mathematical and Computer Modeling}, 16(8/9):27--40, 1992.

\bibitem[CKS81]{cha-koz-sto:j:alternation}
A.~Chandra, D.~Kozen, and L.~Stockmeyer.
\newblock Alternation.
\newblock {\em Journal of the ACM}, 26(1):114--133, 1981.

\bibitem[CLM{\etalchar{+}}12]{che-lan-mau-mon-xia:j:possible-winners-new-candidates-scoring}
Y.~Chevaleyre, J.~Lang, N.~Maudet, J.~Monnot, and L.~Xia.
\newblock New candidates welcome! {Possible} winners with respect to the
  addition of new candidates.
\newblock {\em Mathematical Social Sciences}, 64(1):74--88, 2012.

\bibitem[CSL07]{con-lan-san:j:when-hard-to-manipulate}
V.~Conitzer, T.~Sandholm, and J.~Lang.
\newblock When are elections with few candidates hard to manipulate?
\newblock {\em Journal of the ACM}, 54(3):Article~14, 2007.

\bibitem[DE10]{des-elk:c:sequential-voting}
Y.~Desmedt and E.~Elkind.
\newblock Equilibria of plurality voting with abstentions.
\newblock In {\em Proceedings of the 11th ACM Conference on Electronic
  Commerce}, pages 347--356. ACM Press, June 2010.

\bibitem[DP01]{dek-pic:j:sequential-voting-binary-elections}
E.~Dekel and M.~Piccione.
\newblock Sequential voting procedures in symmetric binary elections.
\newblock {\em Journal of Political Economy}, 108(1):34--55, 2001.

\bibitem[EFRS15]{erd-fel-rot-sch:j:control-in-bucklin-and-fallback-voting}
G.~Erd\'{e}lyi, M.~Fellows, J.~Rothe, and L.~Schend.
\newblock Control complexity in {Bucklin} and fallback voting: {A} theoretical
  analysis.
\newblock {\em Journal of Computer and System Sciences}, 81(4):632--660, 2015.

\bibitem[EHH14]{erd-hem-hem:c:vru-bribery}
G.~Erd\'{e}lyi, E.~Hemaspaandra, and L.~Hemaspaandra.
\newblock Bribery and voter control under voting-rule uncertainty.
\newblock In {\em Proceedings of the 13th International Conference on
  Autonomous Agents and Multiagent Systems}, pages 61--68. International
  Foundation for Autonomous Agents and Multiagent Systems, May 2014.

\bibitem[EHH15]{erd-hem-hem:c:more-natural-models-of-partition-control}
G.~Erd\'{e}lyi, E.~Hemaspaandra, and L.~Hemaspaandra.
\newblock More natural models of electoral control by partition.
\newblock In {\em Proceedings of the 4th International Conference on
  Algorithmic Decision Theory}, pages 396--413. Springer-Verlag {\it Lecture
  Notes in Artificial Intelligence \#9346}, September 2015.

\bibitem[ENR09]{erd-now-rot:j:sp-av}
G.~Erd\'{e}lyi, M.~Nowak, and J.~Rothe.
\newblock Sincere-strategy preference-based approval voting fully resists
  constructive control and broadly resists destructive control.
\newblock {\em Mathematical Logic Quarterly}, 55(4):425--443, 2009.

\bibitem[FHH11]{fal-hem-hem:j:multiprong}
P.~Faliszewski, E.~Hemaspaandra, and L.~Hemaspaandra.
\newblock Multimode control attacks on elections.
\newblock {\em Journal of Artificial Intelligence Research}, 40:305--351, 2011.

\bibitem[FHH13]{fit-hem-hem:c:control-manipulation}
Z.~Fitzsimmons, E.~Hemaspaandra, and L.~Hemaspaandra.
\newblock Control in the presence of manipulators: {Cooperative} and
  competitive cases.
\newblock In {\em Proceedings of the 23rd International Joint Conference on
  Artificial Intelligence}, pages 113--119. AAAI Press, August 2013.

\bibitem[FHH14]{fal-hem-hem:j:single-peaked-nearly}
P.~Faliszewski, E.~Hemaspaandra, and L.~Hemaspaandra.
\newblock The complexity of manipulative attacks in nearly single-peaked
  electorates.
\newblock {\em Artificial Intelligence}, 207:69--99, 2014.

\bibitem[FHHR09]{fal-hem-hem-rot:j:llull}
P.~Faliszewski, E.~Hemaspaandra, L.~Hemaspaandra, and J.~Rothe.
\newblock Llull and {Copeland} voting computationally resist bribery and
  constructive control.
\newblock {\em Journal of Artificial Intelligence Research}, 35:275--341, 2009.

\bibitem[FHHR11]{fal-hem-hem-rot:j:single-peaked-control}
P.~Faliszewski, E.~Hemaspaandra, L.~Hemaspaandra, and J.~Rothe.
\newblock The shield that never was: {Societies} with single-peaked preferences
  are more open to manipulation and control.
\newblock {\em Information and Computation}, 209(2):89--107, 2011.

\bibitem[FR16]{fal-rot:b:handbook-comsoc-control-and-bribery}
P.~Faliszewski and J.~Rothe.
\newblock Control and bribery in voting.
\newblock In F.~Brandt, V.~Conitzer, U.~Endriss, J.~Lang, and A.~Procaccia,
  editors, {\em Handbook of Computational Social Choice}, pages 146--168.
  Cambridge University Press, 2016.

\bibitem[HHR07]{hem-hem-rot:j:destructive-control}
E.~Hemaspaandra, L.~Hemaspaandra, and J.~Rothe.
\newblock Anyone but him: {The} complexity of precluding an alternative.
\newblock {\em Artificial Intelligence}, 171(5--6):255--285, 2007.

\bibitem[HHR12a]{hem-hem-rot:c:online-candidate-control}
E.~Hemaspaandra, L.~Hemaspaandra, and J.~Rothe.
\newblock Controlling candidate-sequential elections.
\newblock In {\em Proceedings of the 20th European Conference on Artificial
  Intelligence}, pages 905--906. IOS Press, August 2012.

\bibitem[HHR12b]{hem-hem-rot:c:online-voter-control}
E.~Hemaspaandra, L.~Hemaspaandra, and J.~Rothe.
\newblock Online voter control in sequential elections.
\newblock In {\em Proceedings of the 20th European Conference on Artificial
  Intelligence}, pages 396--401. IOS Press, August 2012.

\bibitem[HHR14]{hem-hem-rot:j:online-manipulation}
E.~Hemaspaandra, L.~Hemaspaandra, and J.~Rothe.
\newblock The complexity of online manipulation of sequential elections.
\newblock {\em Journal of Computer and System Sciences}, 80(4):697--710, 2014.

\bibitem[Men13]{men:j:range-voting}
C.~Menton.
\newblock Normalized range voting broadly resists control.
\newblock {\em Theory of Computing Systems}, 53(4):507--531, 2013.

\bibitem[OL14]{luc-ore:c:budgeted-social-choice}
J.~Oren and B.~Lucier.
\newblock Online (budgeted) social choice.
\newblock In {\em Proceedings of the 28th AAAI Conference on Artificial
  Intelligence}, pages 1456--1462. AAAI Press, July 2014.

\bibitem[PP13]{par-pro:c:dynamic-social-choice}
D.~Parkes and A.~Procaccia.
\newblock Dynamic social choice with evolving preferences.
\newblock In {\em Proceedings of the 27th AAAI Conference on Artificial
  Intelligence}, pages 767--773. AAAI Press, July 2013.

\bibitem[RS13]{rot-sch:j:typical-case-challenges}
J.~Rothe and L.~Schend.
\newblock Challenges to complexity shields that are supposed to protect
  elections against manipulation and control: {A} survey.
\newblock {\em Annals of Mathematics and Artificial Intelligence},
  68(1--3):161--193, 2013.

\bibitem[Slo93]{slo:j:sequential-voting}
B.~Sloth.
\newblock The theory of voting and equilibria in noncooperative games.
\newblock {\em Games and Economic Behavior}, 5(1):152--169, 1993.

\bibitem[Ten04]{ten:c:transitive-voting}
M.~Tennenholtz.
\newblock Transitive voting.
\newblock In {\em Proceedings of the 5th ACM Conference on Electronic
  Commerce}, pages 230--231. ACM Press, July 2004.

\bibitem[XC10]{con-xia:c:stackelberg-sequential}
L.~Xia and V.~Conitzer.
\newblock Stackelberg voting games: {Computational} aspects and paradoxes.
\newblock In {\em Proceedings of the 24th AAAI Conference on Artificial
  Intelligence}, pages 697--702. AAAI Press, July 2010.

\end{thebibliography}
\newcommand{\etalchar}[1]{$^{#1}$}

\end{document}